\documentclass{PoS}

\title{Z+jet production at NNLO}

\ShortTitle{Z+jet production at NNLO}

\author{A.\ Gehrmann--De Ridder\\
        Institute for Theoretical Physics, ETH, CH-8093 Z\"urich, Switzerland\\
        Department of Physics, University of Z\"urich, CH-8057 Z\"urich, Switzerland\\
        Kavli Institute for Theoretical Physics, UC Santa Barbara, Santa Barbara, USA\\
        E-mail: \email{gehra@phys.ethz.ch}}

\author{T.\ Gehrmann\\
        Department of Physics, University of Z\"urich, CH-8057 Z\"urich, Switzerland\\
        Kavli Institute for Theoretical Physics, UC Santa Barbara, Santa Barbara, USA\\
        E-mail: \email{thomas.gehrmann@uzh.ch}}

\author{E.W.N.\ Glover\\
        Institute for Particle Physics Phenomenology, Department of Physics, University of Durham, Durham, DH1 3LE, UK\\
        E-mail: \email{e.w.n.glover@durham.ac.uk}}

\author{\speaker{A.\ Huss}\\
      Institute for Theoretical Physics, ETH, CH-8093 Z\"urich, Switzerland\\
      E-mail: \email{ahuss@phys.ethz.ch}}

\author{T.A.\ Morgan\\
        Institute for Particle Physics Phenomenology, Department of Physics, University of Durham, Durham, DH1 3LE, UK\\
        E-mail: \email{t.a.morgan@durham.ac.uk}}

\abstract{
We give a brief overview of our calculation of the next-to-next-to-leading order (NNLO) QCD corrections to $\mathrm{Z}+\mathrm{jet}$ production in hadronic collisions.
Phenomenological results are presented which comprise various differential distributions for $8~\mathrm{TeV}$ proton--proton collisions.
A significant reduction of the scale uncertainties is observed throughout as we move from NLO to NNLO.
We further discuss how this calculation can be used to describe the inclusive Z-boson production at large transverse momentum.
To this end, the theory prediction is compared to the measurements performed by the ATLAS and CMS collaborations at a centre-of-mass energy of $8~\mathrm{TeV}$.
Here, the inclusion of NNLO QCD effects are found to result in a substantial improvement in the agreement between theory and data for the normalised distributions.
}

\FullConference{Loops and Legs in Quantum Field Theory\\
		24-29 April 2016\\
		Leipzig, Germany}

\begin{document}

\section{Introduction}

The Drell--Yan-like production of charged lepton pairs, $\mathrm{p}\mathrm{p}\to\mathrm{Z}/\gamma^*\to\ell^+\ell^-$, is among the most important so-called ``standard candle'' processes at the LHC, owing to its large production rate and clean experimental signature. 
It provides the opportunity to test the Standard Model prediction over a large kinematic range and delivers crucial constraints in the fit of parton distribution functions (PDFs). Moreover, it represents a powerful tool for detector calibration and also constitutes an important background to many searches for physics beyond the Standard Model.

The production of a Z boson in association with a hadronic jet still retains a large event rate and also has the advantage of introducing a direct sensitivity to the strong coupling $\alpha_\mathrm{s}$ and the gluon PDF. As such, $\mathrm{Z}+\mathrm{jet}$ production provides an ideal testing ground for our understanding of both strong and electroweak interactions in a hadron-collider environment~\cite{ATLASZJ,CMSZJ}.

On the theory side, a variety of corrections have been considered in order to improve the accuracy of the theoretical predictions for the Z+jet process. Both the NLO QCD~\cite{ZJNLO} and EW~\cite{ZJNLOEW} corrections to this process have been known for some time. 
The NLO multi-jet merging procedure has been recently extended to incorporate both NLO QCD and EW corrections into the \textsc{MePs@Nlo} framework~\cite{ZJNLOEWM}. 
QCD corrections to Z+jet production at NNLO were computed in Refs.~\cite{ZJNNLOus,ZJBoughezal} using two independent methods and have subsequently been validated against each other.

\section{Z-boson production in association with a hadronic jet}

The calculation of the NNLO corrections to $\mathrm{Z}+\mathrm{jet}$ production presented in Ref.~\cite{ZJNNLOus} employs the antenna-subtraction formalism~\cite{ourant}, which redistributes infrared singularities among the contributions of different parton multiplicities via the introduction of local subtraction terms.
All the basic building blocks of this subtraction method are known analytically for the configurations that are relevant for hadron--hadron collisions.
The results are implemented in a flexible parton-level Monte Carlo program \textsc{NNLOjet} which allows the computation of cross sections and any (multi-) differential distribution with arbitrary (infrared-safe) event-selection cuts at NNLO.

\begin{figure}[t]
  \includegraphics[angle=0,width=.5\linewidth]{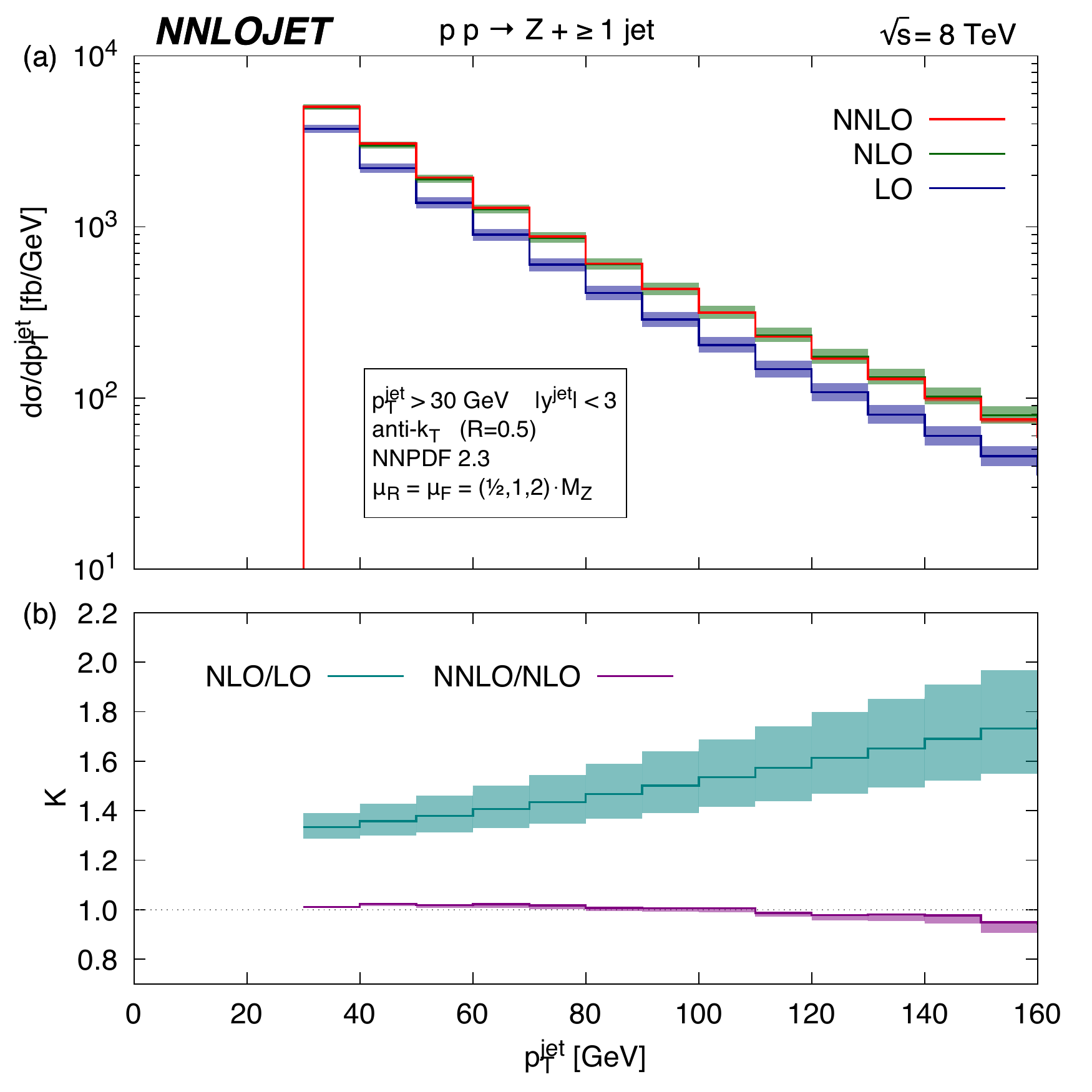}
  \hfill
  \includegraphics[angle=0,width=.5\linewidth]{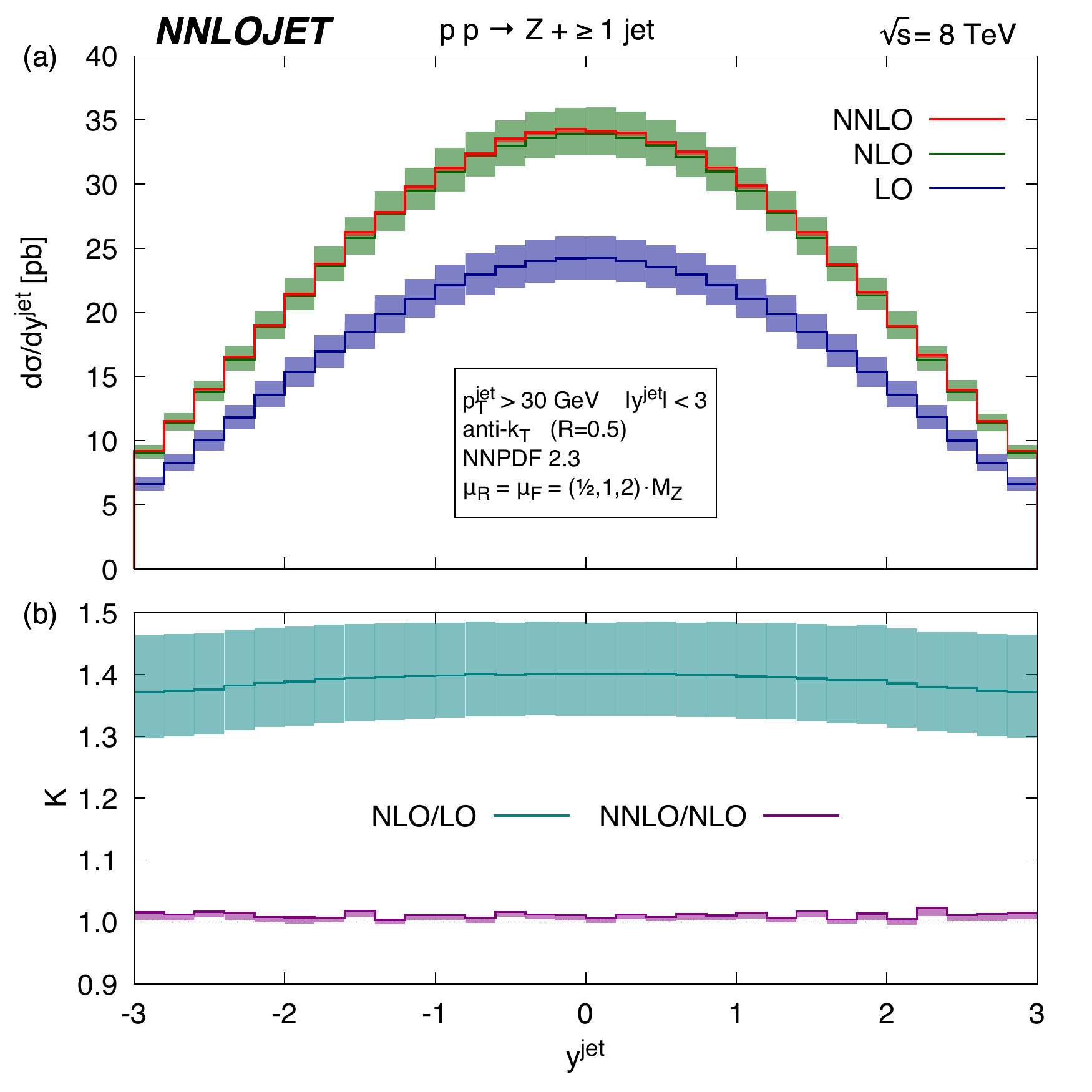}
  \caption{The transverse momentum (left) and rapidity (right) distribution of the leading jet in inclusive $Z+\mathrm{jet}$ production in $pp$ collisions with 
  $\sqrt{s}=8~\mathrm{TeV}$.
  The absolute distributions at LO (blue), NLO (green), and NNLO (red) are shown in the top panels.
  The bottom frames display the ratios of different perturbative orders: NLO to LO (turquoise) and NNLO to NLO (mauve).}
  \label{figs:j1}
\end{figure}

\begin{figure}[t]
  \includegraphics[angle=0,width=.5\linewidth]{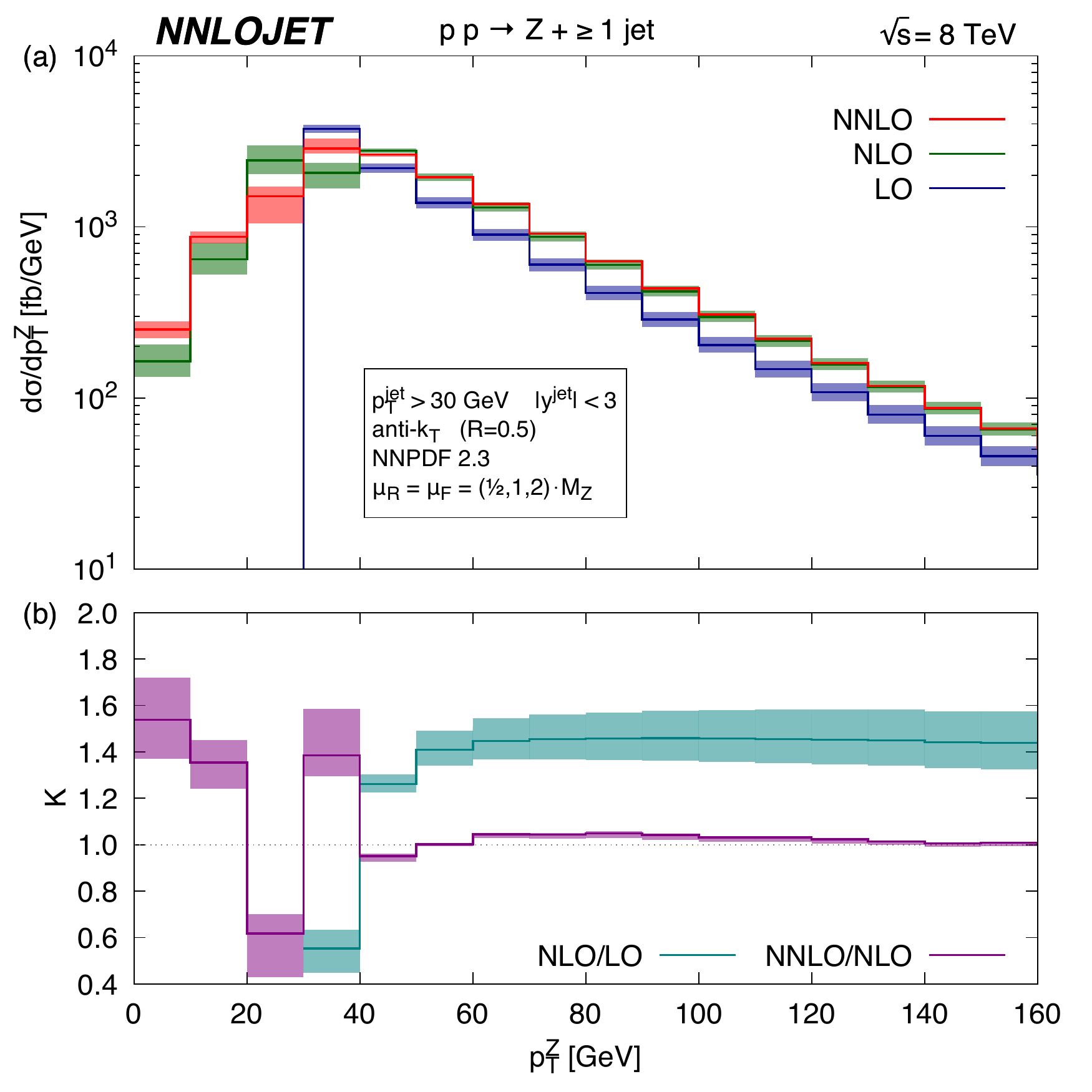}
  \hfill
  \includegraphics[angle=0,width=.5\linewidth]{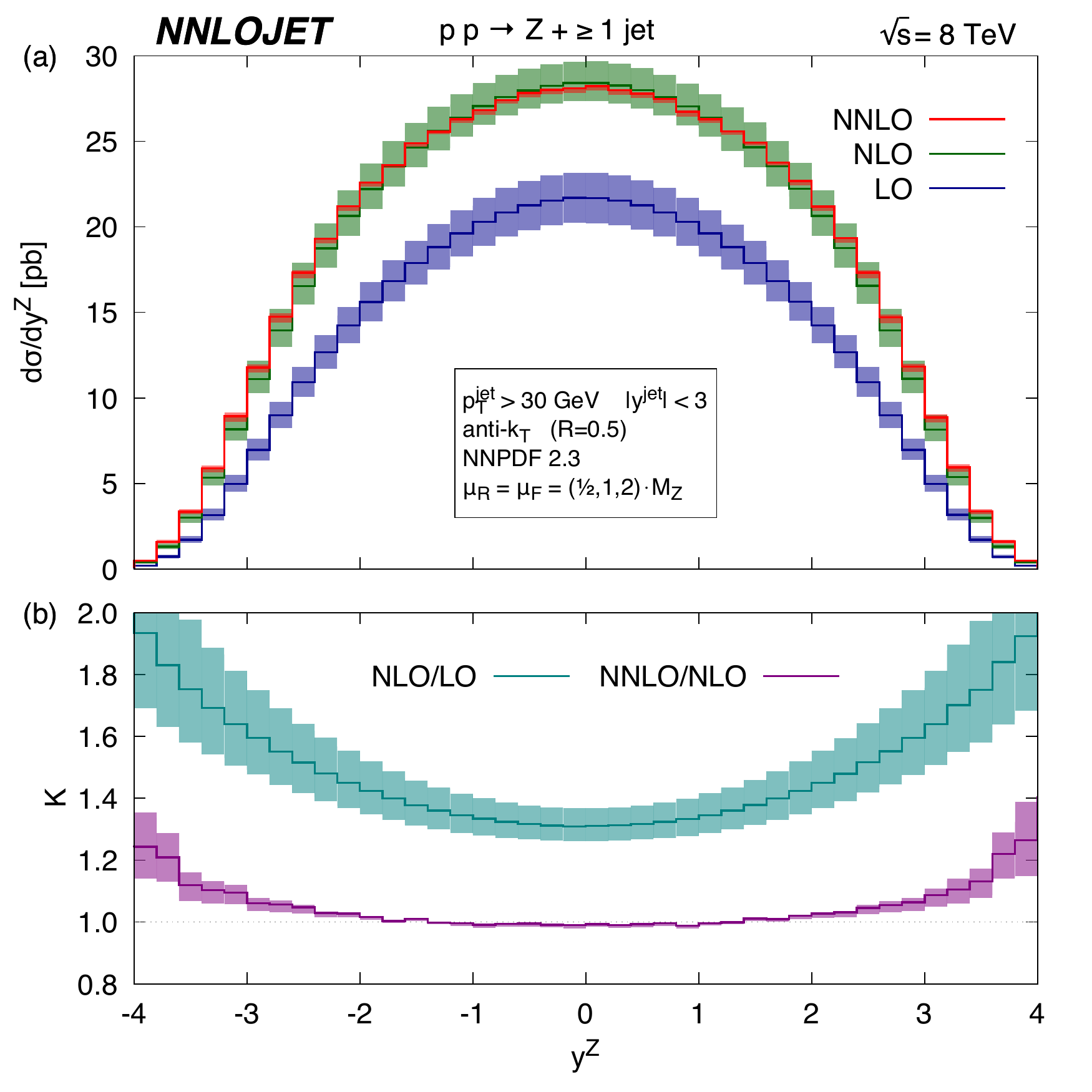}
  \caption{The transverse momentum (left) and rapidity (right) distribution of the Z boson in inclusive $Z+\mathrm{jet}$ production in $pp$ collisions with 
  $\sqrt{s}=8~\mathrm{TeV}$.
  The absolute distributions at LO (blue), NLO (green), and NNLO (red) are shown in the top panels.
  The bottom frames display the ratios of different perturbative orders: NLO to LO (turquoise) and NNLO to NLO (mauve).}
  \label{figs:Z}
\end{figure}

In Figs.~\ref{figs:j1} and \ref{figs:Z}, we present the numerical results for the leading-jet and Z-boson observables, respectively.
The top frames~(a) of the plots display the absolute distributions for the different perturbative orders, while the the relative impact of the (N)NLO corrections are exposed in the bottom panels~(b) in terms of the ratio $K = {\mathrm{d}\sigma^\mathrm{(N)NLO}(\mu)}/{\mathrm{d}\sigma^\mathrm{(N)LO}(\mu=M_\mathrm{Z})}$.
The theoretical uncertainties shown as bands around the curves are estimated by varying the unphysical scales $\mu\equiv\mu_\mathrm{R}=\mu_\mathrm{F}$ by the factors $[1/2,2]$ from the central scale choice $\mu=M_\mathrm{Z}$.
For more results and details of the numerical setup, we refer the reader to Refs.~\cite{ZJNNLOus,radcor}.

The left plot in Fig.~\ref{figs:j1} shows the leading-jet transverse momentum distribution, where we observe NLO corrections that are at the level of $30$--$70\%$ with a residual scale uncertainty of $5$--$10\%$.
This uncertainty is significantly reduced as we move from NLO to NNLO and we further observe a stabilisation of the perturbative series with NNLO corrections that amount to less than an additional $5\%$ correction.
The rapidity distribution of the leading jet is shown in the right plot of Fig.~\ref{figs:j1}. The NLO and NNLO corrections are relatively flat in this distribution and amount to $35$--$40\%$ and $\sim1\%$, respectively.
We again observe a considerable reduction of the scale uncertainty as we move from NLO to NNLO.

Figure~\ref{figs:Z} shows the transverse momentum~(left) and rapidity~(right) distribution of the $\mathrm{Z}$ boson.
An interesting structure around $p_\mathrm{T}^\mathrm{Z}\approx 30~\mathrm{GeV}$ is observed in the transverse momentum distribution.
This is the well-known Sudakov shoulder phenomenon~\cite{sudakov}, where the indirect constraint on the leading-order process, $p_\mathrm{T}^\mathrm{Z}>30~\mathrm{GeV}$, is alleviated by real-radiation corrections at higher orders.
At larger transverse momenta, the NNLO corrections increase the prediction by approximately $1\%$. 
The rapidity of the Z boson is shown in the right plot of Fig.~\ref{figs:Z}.
The NLO and NNLO corrections are the largest in the forward/backward region where they can reach corrections of up to $\sim90\%$ and $\sim20\%$, respectively.
In the central region, the NNLO corrections are very small with a reduced scale dependence.

In the differential distributions,  we observe that the corrections are not always uniform. This implies that a rescaling of lower-order predictions is insufficient for precision applications.

\section{Inclusive Z-boson production at high transverse momentum}

A particularly interesting observable is the transverse momentum of the Z boson, which is caused by partonic recoil and is thus determined by QCD dynamics.
One of the main motivations to study this observable, in particular multi-differentially, is its sensitivity to the gluon distribution which is still poorly constrained at high $x$ values. 
It is important to note that predictions which are accurate to NNLO in QCD for the inclusive Z-boson production cross section are only NLO-accurate in this observable due to the aforementioned partonic recoil.
ATLAS and CMS both observed a tension between their measurements and existing NLO QCD predictions, highlighting the potential importance of higher order corrections to this process.

Building upon our calculation of the NNLO QCD corrections to $\mathrm{Z}+\mathrm{jet}$ discussed in the previous section, we exploit our highly flexible \textsc{NNLOjet} numerical code  to predict the Z-boson transverse momentum distribution to NNLO accuracy.
To this end, we now consider a setup that is fully inclusive with respect to QCD radiation and instead impose a small transverse momentum cut on the Z boson, enforcing the presence of a non-vanishing hadronic recoil.

The experimental measurement of the transverse momentum of the Z boson, $p_\mathrm{T}^\mathrm{Z}$, is presented in the form of fiducial cross sections for a restricted kinematical range of the final state leptons. 
We compare our NNLO calculation to data by considering the same event selection cuts as used in the ATLAS~\cite{ptzATLAS} and CMS~\cite{ptzCMS} analyses based on data collected in Run I of the LHC with $\sqrt{s}=8~\mathrm{TeV}$.
In order to reduce the systematic uncertainty on the measurement, in particular the luminosity uncertainty of about $3\%$, the transverse momentum distribution is commonly normalised to the inclusive production cross section,
\begin{equation}
  \frac{1}{\sigma} \cdot \frac{\mathrm{d}\sigma}{\mathrm{d}p_\mathrm{T}^\mathrm{Z}}.
  \label{eq:ptznorm}
\end{equation}
We compute the NNLO corrections to the unnormalised $p_\mathrm{T}^\mathrm{Z}$ distribution using our recent calculation for $\mathrm{Z}+\mathrm{jet}$ production by imposing a transverse momentum cut of $20~\mathrm{GeV}$
\begin{equation}
  \frac{\mathrm{d}\sigma}{\mathrm{d}p_\mathrm{T}^\mathrm{Z}} \;\bigg\vert_{p_\mathrm{T}^\mathrm{Z}>20~\mathrm{GeV}} = 
  \bigg(
  \frac{\mathrm{d}\sigma^{\mathrm{Z}+\mathrm{jet}}_\mathrm{LO}}{\mathrm{d}p_\mathrm{T}^\mathrm{Z}} +
  \frac{\mathrm{d}\sigma^{\mathrm{Z}+\mathrm{jet}}_\mathrm{NLO}}{\mathrm{d}p_\mathrm{T}^\mathrm{Z}} +
  \frac{\mathrm{d}\sigma^{\mathrm{Z}+\mathrm{jet}}_\mathrm{NNLO}}{\mathrm{d}p_\mathrm{T}^\mathrm{Z}} 
  \bigg) \bigg\vert_{p_\mathrm{T}^\mathrm{Z}>20~\mathrm{GeV}}
  + \mathcal{O}(\alpha_\mathrm{s}^4) , 
  \label{eq:ptzabs}
\end{equation}
while the normalisation in Eq.~(\ref{eq:ptznorm}) is obtained from the Drell--Yan cross section
\begin{equation}
  \sigma = \int_0^\infty \frac{\mathrm{d}\sigma}{\mathrm{d}p_\mathrm{T}^\mathrm{Z}} \; \mathrm{d}p_\mathrm{T}^\mathrm{Z} =
  \sigma^\mathrm{Z}_\mathrm{LO} + \sigma^\mathrm{Z}_\mathrm{NLO} + \sigma^\mathrm{Z}_\mathrm{NNLO} 
  + \mathcal{O}(\alpha_\mathrm{s}^3) .
  \label{eq:norm}
\end{equation}

\begin{figure}[t]
  \raisebox{-0.5\height}{\includegraphics[width=.5\linewidth]{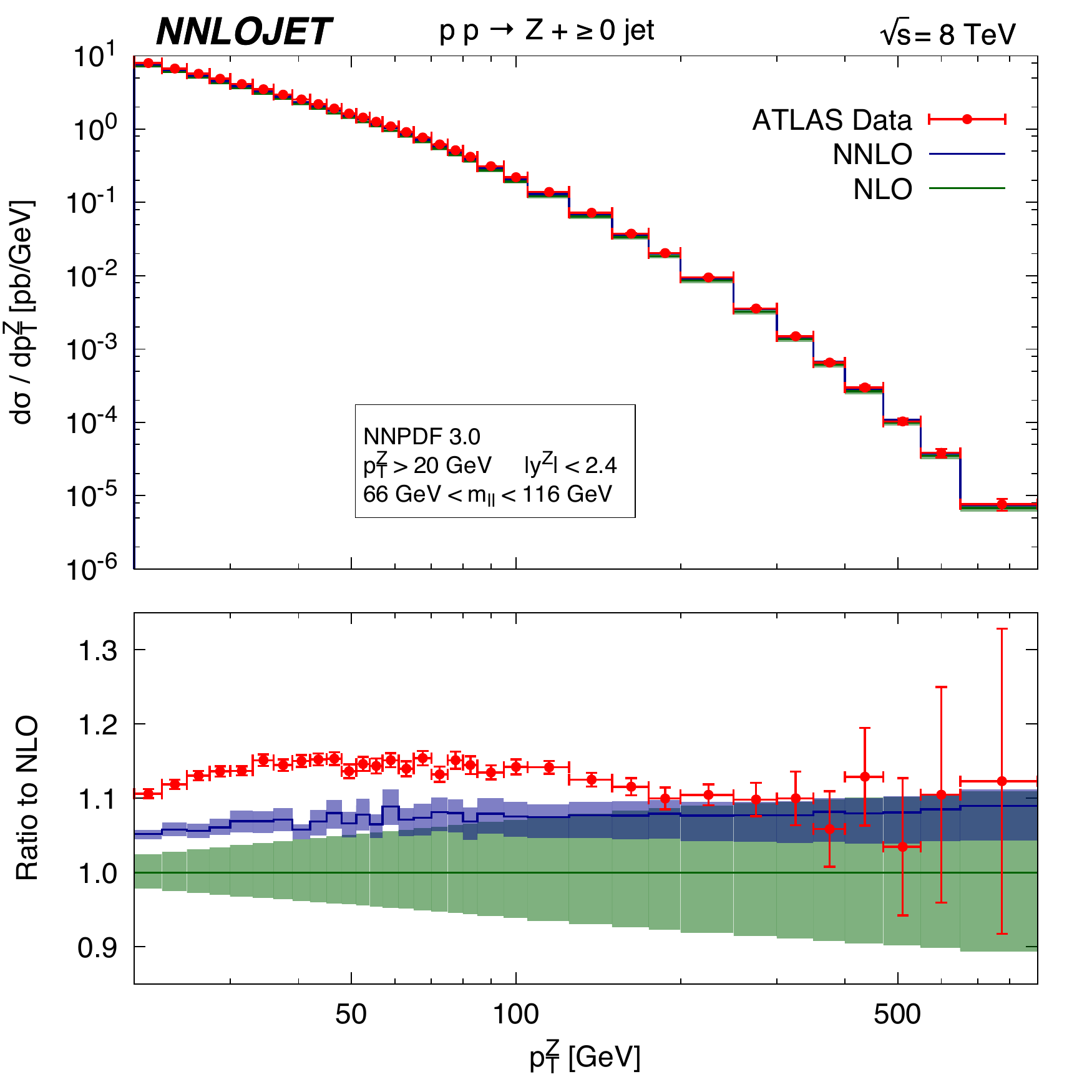}}
  \hfill
  \raisebox{-0.5\height}{\includegraphics[width=.5\linewidth]{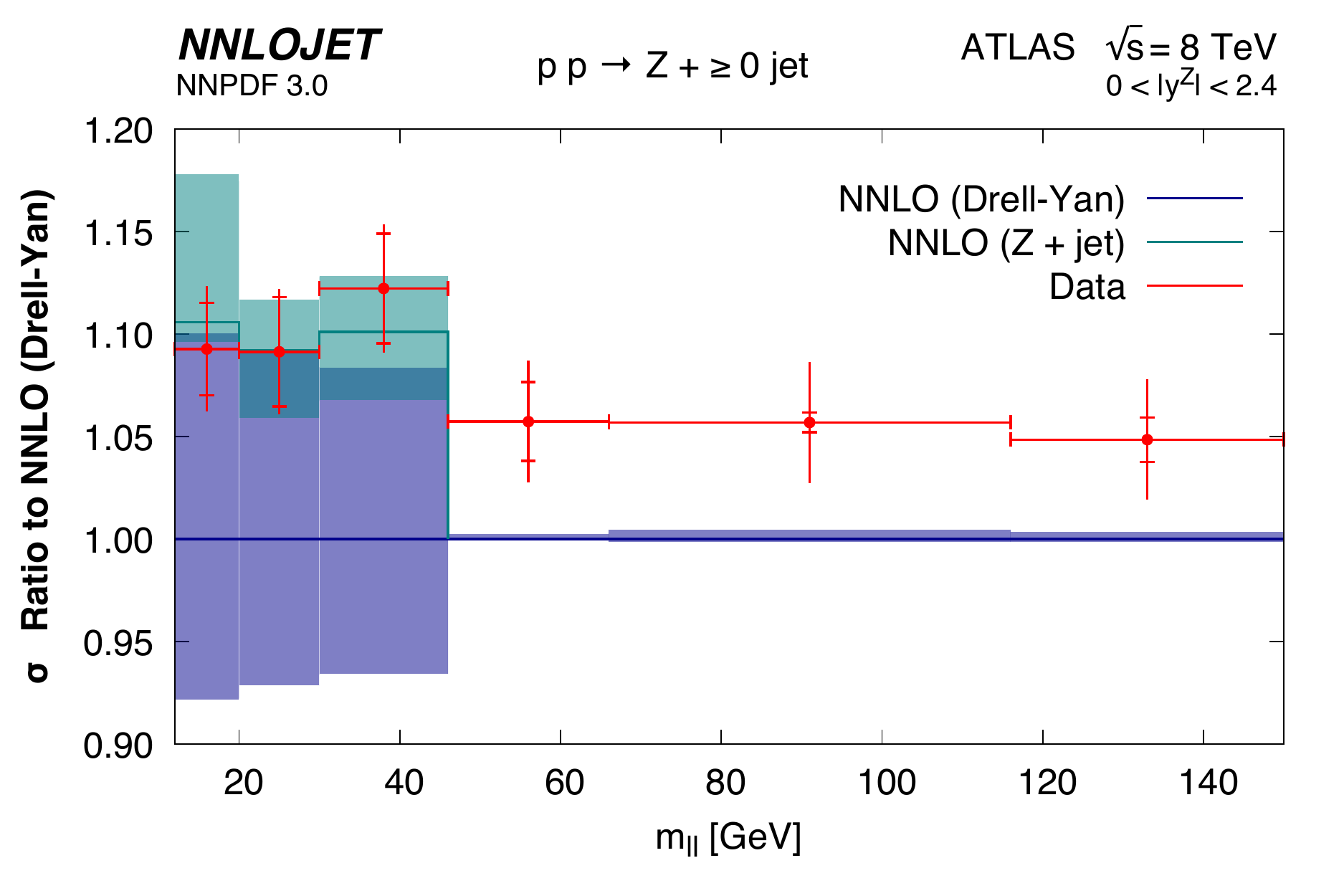}}
  \caption{
    Left: The unnormalised Z-boson transverse momentum distribution for the $66~\mathrm{GeV}<m_{\ell\ell}<116~\mathrm{GeV}$ bin of the ATLAS measurement. 
    The green/blue bands denote the NLO/NNLO prediction with scale uncertainty.
    Right: The inclusive Drell--Yan cross section for different $m_{\ell\ell}$ bins used in the ATLAS analysis.
  }
  \label{figs:norm}
\end{figure}
As discussed in Ref.~\cite{PTZus}, the inclusion of the NNLO QCD effects does not fully resolve the tension with the data for the unnormalised $p_\mathrm{T}^\mathrm{Z}$ distribution.
This is illustrated in the left plot of Fig.~\ref{figs:norm}, where the fixed-order predictions are compared to the ATLAS data in the $66~\mathrm{GeV}<m_{\ell\ell}<116~\mathrm{GeV}$ bin.
In the bottom panel, we display the ratios with respect to the NLO prediction where we can observe that the NNLO corrections increase the cross section by about $5\%$.
The bands on the theory curves represent the scale uncertainty as obtained through a variation of the renormalisation and factorisation scale by a factor in the range $[1/2, 2]$ from the central scale choice of $\mu=\sqrt{m_{\ell\ell}^2+(p_\mathrm{T}^\mathrm{Z})^2}$ and are greatly reduced by moving to NNLO from NLO.
Despite the observed shift of the theory prediction towards the data points, a systematic offset of about $5\%$ remains in the data--theory comparison which cannot be fully accounted for by the overall $2.8\%$ luminosity error that is not included in the data points shown in Fig.~\ref{figs:norm}~(left). 
However, inspecting the Drell--Yan fiducial cross section, i.e.\ the normalisation in Eq.~(\ref{eq:norm}), reveals a systematically larger value for the measured cross section compared to the NNLO prediction.
The difference, as shown in the right-hand plot of Fig.~\ref{figs:norm}, is comparable in size to the data--theory offset found in the unnormalised distributions.
Indeed, considering normalised distributions according to Eq.~(\ref{eq:ptznorm}) leads to a substantial improvement in the agreement between theory and data as will be discussed below.

The three lowest mass bins in Fig.~\ref{figs:norm}~(right) display a much larger scale uncertainty compared to the higher mass bins as a consequence of the additional event selection cut, $p_\mathrm{T}^\mathrm{Z}>45~\mathrm{GeV}$, that forbids the low-mass bins to be populated at LO for the Drell--Yan process.
Therefore, our NNLO prediction for the normalisation in Eq.~(\ref{eq:norm}) is effectively only NLO accurate in these bins, with consequently larger scale dependence.
We can improve the theory prediction to genuine NNLO-accuracy by using the $\mathrm{Z}+\mathrm{jet}$ inclusive cross section at NNLO with $p_\mathrm{T}^\mathrm{Z}>45~\mathrm{GeV}$,
\begin{equation}
  \sigma \;\big\vert_{p_\mathrm{T}^\mathrm{Z}>45~\mathrm{GeV}} = 
  \int_{45~\mathrm{GeV}}^\infty \frac{\mathrm{d}\sigma}{\mathrm{d}p_\mathrm{T}^\mathrm{Z}} \; \mathrm{d}p_\mathrm{T}^\mathrm{Z} =
  \bigl( \sigma^{\mathrm{Z}+\mathrm{jet}}_\mathrm{LO} + \sigma^{\mathrm{Z}+\mathrm{jet}}_\mathrm{NLO} + \sigma^{\mathrm{Z}+\mathrm{jet}}_\mathrm{NNLO} 
  \bigr)\bigr\vert_{p_\mathrm{T}^\mathrm{Z}>45~\mathrm{GeV}}
  + \mathcal{O}(\alpha_\mathrm{s}^4) .
  \label{eq:norm2}
\end{equation}
The result for the fiducial cross section obtained using Eq.~(\ref{eq:norm2}) are shown as the turquoise curve in Fig.~\ref{figs:norm}~(right).
The scale uncertainty is reduced by more than a factor of two compared to the only NLO-accurate prediction and, moreover, we observe a shift of the central value towards the measured cross section values. 
In the remainder of this section we will restrict ourselves to the discussion of normalised distributions where we use the improved normalisation of Eq.~(\ref{eq:norm2}) for the three lowest-mass bins of the ATLAS measurement.
For the respective comparison of unnormalised distributions and further results we refer to Ref.~\cite{PTZus}, where also more details are given.

\begin{figure}[t]
  \centering
  \includegraphics[width=.9\linewidth]{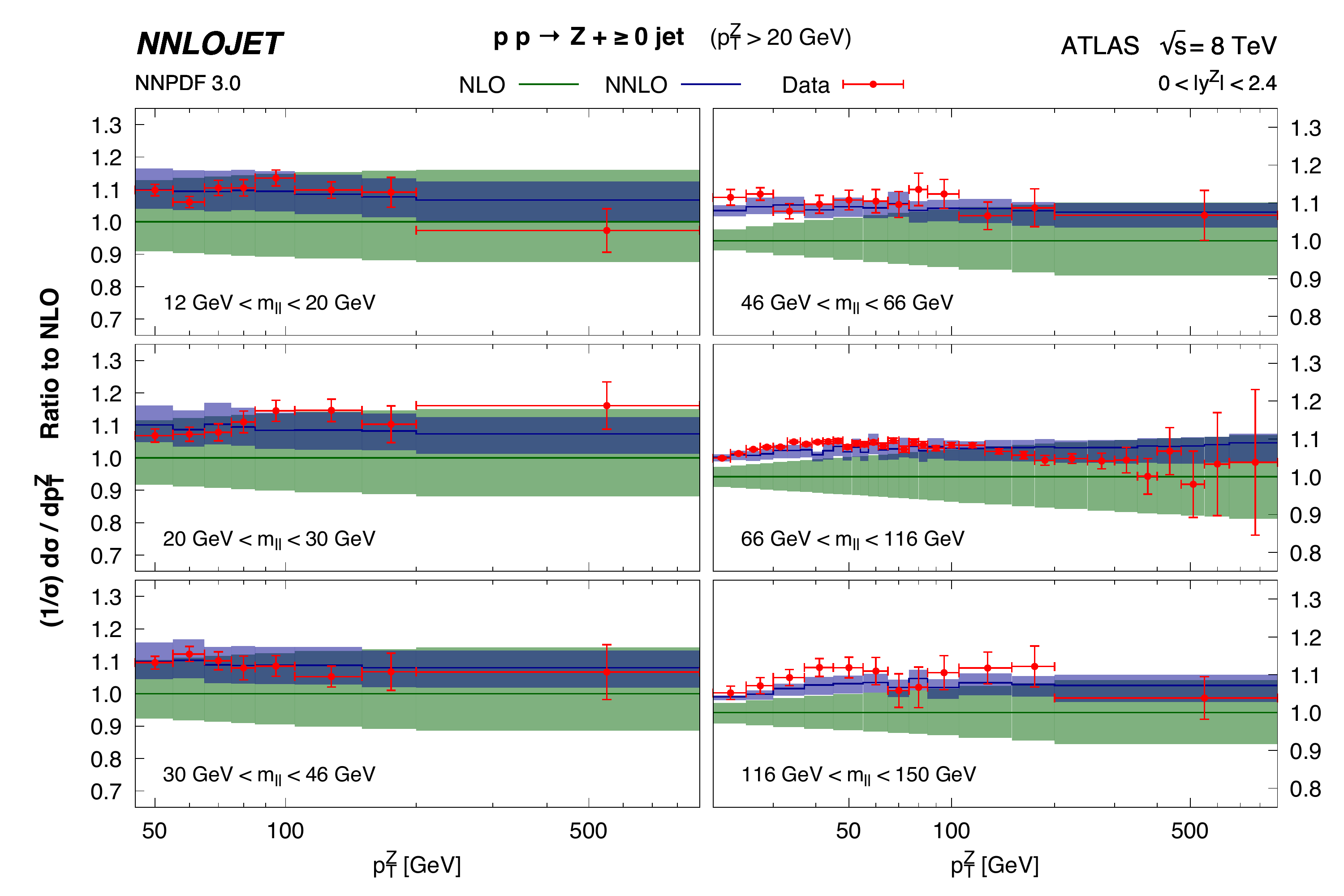}
  \caption[]{The normalised double-differential transverse momentum distribution for the Z boson in windows of invariant mass of the leptons, $m_{\ell\ell}$, with a rapidity cut on the Z boson of $|y^\mathrm{Z}| < 2.4$. The ATLAS data is taken from~\cite{ptzATLAS}.  The green bands denote the NLO prediction with scale uncertainty and the blue bands show the NNLO prediction with scale uncertainty.}
  \label{fig:normATLASmll}
\end{figure}
In Fig.~\ref{fig:normATLASmll} we present the normalised double-differential distribution with respect to the transverse momentum of the Z boson and the invariant mass of the lepton pair, $m_{\ell\ell}$, normalised to the NLO prediction and compare it to the ATLAS data~\cite{ptzATLAS}. Tension between the NLO prediction and the data is seen in the three higher mass bins where the data is significantly overshooting the theory prediction. 
The NNLO corrections in these bins are not uniform and have a large positive correction at small $p_\mathrm{T}^\mathrm{Z}$. This is particularly apparent for the $m_{\ell\ell}$ bin containing the Z-boson resonance.%
\footnote{Note that the bin including the $\mathrm{Z}$ resonance given by the middle frame in the right column of Fig.~\ref{fig:normATLASmll} corresponds to the normalised distribution of the associated unnormalised distribution shown in the left plot of Fig.~\ref{figs:norm}.}
The tension observed at NLO is completely resolved by the inclusion of the NNLO corrections and we see very good agreement between data and theory.

\begin{figure}[t]
  \centering
  \includegraphics[width=.9\linewidth]{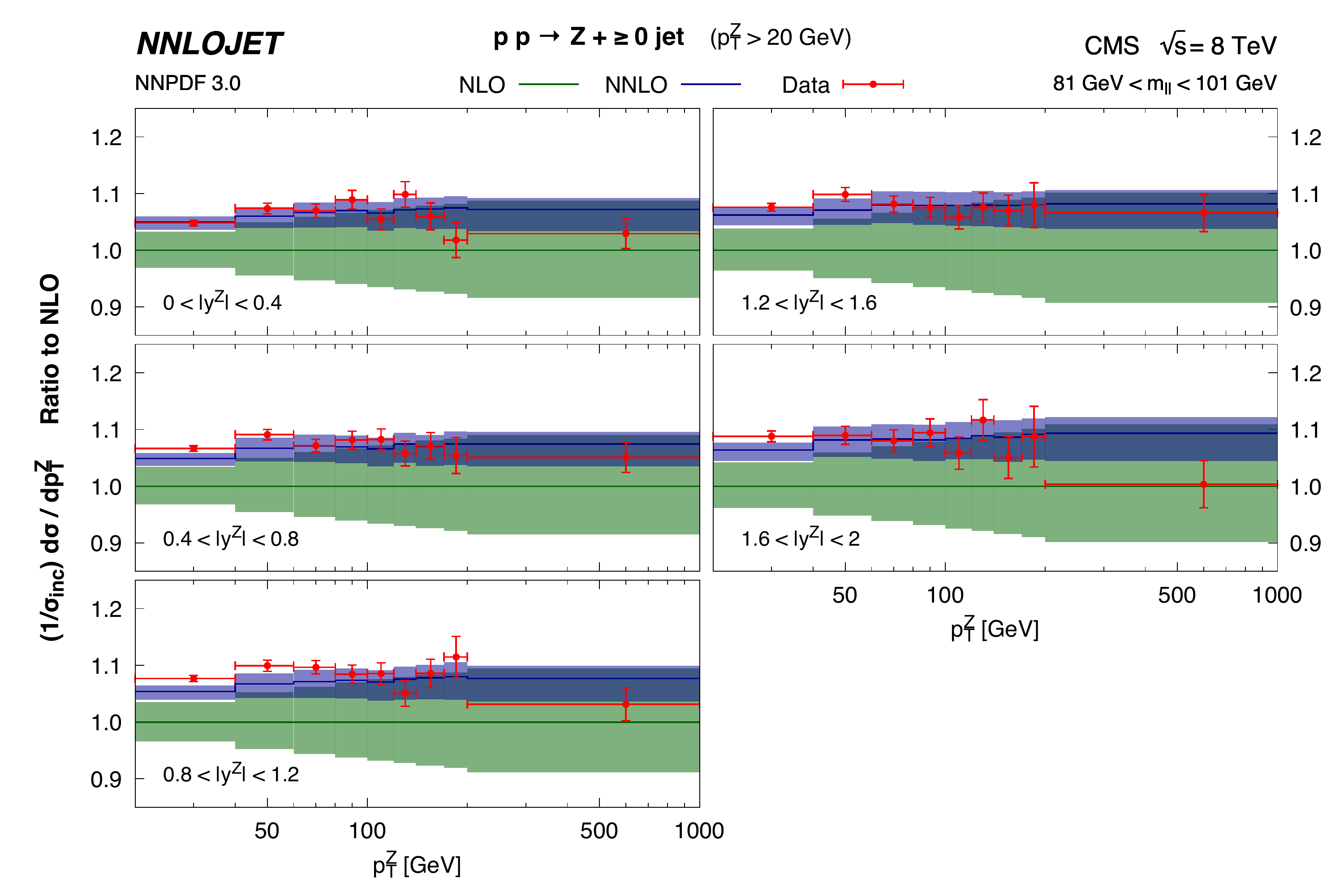}
  \caption[]{The normalised double-differential transverse momentum distribution for the Z boson in windows of rapidity of the Z boson, $y^\mathrm{Z}$, with an invariant mass cut on final state leptons of $81$~GeV~$<m_{\ell\ell} < 101$~GeV. The CMS data is taken from~\cite{ptzCMS}.  The green bands denote the NLO prediction with scale uncertainty and the blue bands show the NNLO prediction.}
  \label{fig:normCMSyz}
\end{figure}
Figure~\ref{fig:normCMSyz} shows the comparison of the CMS measurement~\cite{ptzCMS} to our calculation for the normalised double-differential distribution with respect to $p_\mathrm{T}^\mathrm{Z}$ and the rapidiy of the Z boson ($y^\mathrm{Z}$), as a ratio to the NLO prediction. The tension between the theoretical prediction and the data is similar to the ATLAS analysis~\cite{ptzATLAS}, where the measured data points lie systematically above the prediction.
The NNLO corrections are relatively uniform in rapidity and transverse momentum in the considered kinematic region and amount to about positive $5$--$10\%$ corrections, with a decreased residual theory uncertainty of $2$--$4\%$.
Again, the inclusion of the NNLO corrections completely removes the tension with the data and we obtain excellent agreement.

\section{Conclusions}

We have presented the NNLO QCD corrections to  the $\mathrm{Z}+\mathrm{jet}$ process and to the production of Z-bosons at high transverse momentum.
This calculation is performed using the parton-level Monte Carlo generator \textsc{NNLOjet} which implements the antenna subtraction method for NNLO calculations of hadron collider observables.
We performed a comparison of the theory prediction to the ATLAS~\cite{ptzATLAS} and CMS~\cite{ptzCMS} studies of Z-boson transverse momentum distribution based on the LHC  Run I data.
Although a residual tension between the NNLO QCD theory prediction and experiment remains for the unnormalised inclusive-$p_\mathrm{T}^\mathrm{Z}$ distributions, this is considerably alleviated when the distributions are normalised by the inclusive cross section.

\section*{Acknowledgments}

The authors would like to acknowledge the support provided by the GridPP Collaboration. This research was supported in part by the National Science Foundation under Grant NSF PHY11-25915,
in part by the Swiss National Science Foundation (SNF) under contracts 200020-162487 and CRSII2-160814, in part by the UK Science and Technology Facilities Council, in part by the Research Executive Agency (REA) of the European Union under the Grant Agreement PITN-GA-2012-316704 (``HiggsTools'') and the ERC Advanced Grant MC@NNLO (340983).


\end{document}